\documentstyle[aas2pp4]{article}

\newcommand{\vv}[1]{\mbox{\boldmath $#1$}}    
\newcommand{\vvsmall}[1]{\mbox{\scriptsize\boldmath $#1$}}
\newcommand{\ab}{a_{\rm B}}                   
\newcommand{\am}{a_{\rm m}}                   
\newcommand{\mel}{m_{\rm e}}                  
\newcommand{\mpr}{m_{\rm p}}                  
\newcommand{\mH}{m_{\rm H}}                   
\newcommand{\req}[1]{(\ref{#1})}              
\newcommand{\kp}{K_\perp}                      
\newcommand{\kpvec}{\vv{K}_\perp}

\lefthead{Potekhin \& Pavlov}
\righthead{Photoionization in atmospheres of magnetic neutron stars}

\begin{document}

\title{PHOTOIONIZATION OF HYDROGEN 
IN ATMOSPHERES OF MAGNETIC NEUTRON STARS
}
\author{Alexander Y. Potekhin}
\affil{Ioffe Physical-Technical Institute, 194021,
St.-Petersburg, Russia; palex@astro.ioffe.rssi.ru}
\and
\author{George G. Pavlov}
\affil{Pennsylvania State University, 525 Davey Lab, 
University Park, PA 16802, U.S.A.; pavlov@astro.psu.edu}
\begin{abstract}

The strong magnetic fields ($B\sim 10^{12} - 10^{13}$~G) 
characteristic of neutron stars  make all the 
properties 
of an atom 
strongly dependent on the transverse component $\kpvec$ of
its 
generalized momentum.
In particular, the photoionization process 
is modified substantially: (i) 
threshold energies are decreased as compared with those 
for an atom at rest, (ii) cross section values 
are changed significantly, and (iii)
selection rules valid for atoms at rest are violated by the motion
so that new photoionization channels become allowed.
To calculate the photoionization cross sections,
we, for the first time, employ exact numerical 
treatment of both initial and final atomic states.
This enables us 
to take into account the quasi-bound (autoionizing)
atomic states as well as coupling of different 
ionization channels.
We extend the previous consideration, restricted 
to the so-called centered states corresponding to 
relatively small values of $\kp$,
to arbitrary states of atomic motion.  
We fold the cross sections 
with the thermal distribution of atoms over $\vv{K}$. 
For typical temperatures of 
neutron star atmospheres,
the averaged
cross sections differ substantially from those of atoms at rest.
In particular, the photoionization edges are strongly broadened
by the thermal motion of atoms; this ``magnetic broadening''
exceeds the usual Doppler broadening by orders of magnitude.
The decentered states of the atoms  
give rise to the low-energy component of the photoionization 
cross section. This new component grows significantly 
with increasing temperature 
above $10^{5.5}$ K and decreasing density below 1 g cm$^{-3}$, 
i.~e., for the conditions expected in atmospheres 
of middle-aged neutron stars. 

\end{abstract}


\vspace{5ex}
Accepted for publication in {\it The Astrophysical Journal}

\section{INTRODUCTION}
The thermal-like surface radiation detected recently
from a number of isolated neutron stars (NSs)
(see, e.~g., \"Ogelman 1995 for a review) 
provides valuable information about the NS temperatures and 
radii.
This enables one to trace 
the thermal history of NSs
and to put constraints on 
the properties of the superdense matter 
in NS interiors (e.~g., Pethick 1992). However, the observed spectra 
are substantially modified by reprocessing of the thermal radiation 
in NS atmospheres (Romani 1987). 
Pavlov et al.\ (1995; and references therein) 
have shown that 
the strong magnetic fields of 
NSs, $B\sim 10^{12}-10^{13}$~G, affect the emergent 
spectra significantly. In particular, photoabsorption 
by the strongly magnetized atomic hydrogen gas which presumably covers
the  NS surface should be taken into account. 
The magnetic field strongly affects the structure and radiative
transitions of the hydrogen atom
when the electron cyclotron energy $\hbar\omega_c=
\hbar eB/(\mel c)$ 
is much greater than the Rydberg energy
${\rm Ry} = \mel e^4/(2\hbar^2) = 13.6$~eV, i.~e.,  
$\gamma=\hbar\omega_c/(2{\rm Ry})=B/(2.35\times 10^9~{\rm G})\gg 1$.
In particular, the atom 
becomes stretched along the magnetic field,
the ionization potential is increased 
by a factor of $\sim \ln^2\gamma$, and radiative
transitions become strongly dependent on the photon polarization.

Photoionization of strongly magnetized hydrogen atoms
{\it at rest} has been considered by a number of authors.
The most thorough consideration in the {\em adiabatic approximation},
applicable for very large $\gamma$ and $\omega \ll \omega_c$
($\omega$ is the photon frequency), was 
presented by Potekhin \& Pavlov (1993; hereafter PP93).
This approximation neglects coupling of states with different
values of the Landau quantum number $n$ which specifies
the energy of the electron motion
transverse to the magnetic field (the other quantum number
associated with the transverse motion, the projection
$s$ of the angular momentum onto the field, is exact for
atoms at rest).
Potekhin, Pavlov, \& Ventura (1996; hereafter PPV97) went
beyond the adiabatic approximation and resolved, in particular,
a longstanding contradiction: some authors (e.~g.,  
Schmitt et al.\ 1981; 
Wunner et al.\ 1983; Ventura et al.\ 1992) concluded
that the photoionization cross section is
exactly zero for photons polarized
transversely to the magnetic field, whereas other
authors (e.~g., Gnedin, Pavlov, \& Tsygan 1973;
Miller \& Neuhauser 1991; PP93) obtained finite cross sections.
PP93 showed that the discrepancy rooted in different
forms of the interaction potential (so-called ``velocity form''
and ``length form'') 
employed by the two groups
of authors, and PPV97 proved that the velocity form 
used by the former group resulted in
missing the main contribution to the cross section.
The nonadiabatic approach enabled PPV97 to study 
correlation of different ionization channels and
manifestation of autoionization (Beutler--Fano)
resonances in the photoabsorption spectra.

The above-cited papers did not take into account 
the atomic 
motion inevitable in any real situation. If there were no
strong magnetic field, the motion would only lead to
the trivial Doppler broadening of the spectral lines and
photoionization edges. On the contrary, the finiteness of
the atomic velocities in strong magnetic fields leads
to many qualitatively new effects (Pavlov \& M\'esz\'aros\ 1993;
hereafter PM93) associated with strong distortion of the
atomic structure due the Lorentz electric field generated
by the motion. Since the moving atom loses its cylindrical
symmetry, the projection $s$ of the angular
momentum is no longer a good 
quantum number (in other words, the $s$-adiabatic approach which
neglects coupling of states with different $s$ becomes invalid).
Hence, the selection rules for the
radiative transitions are changed so that new types of transitions become
allowed. Since the atomic structure depends on the 
velocity of the atom, and different atomic states are affected
by the motion differently, a new type of broadening of 
spectral features, {\em magnetic broadening},
becomes the most important.

The energies and
wave functions of the moving atom depend on the transverse component
$\kp$ of a {\it generalized momentum} $\vv{K}$ (see below),
which, contrary to the
canonical and kinetic momenta, is conserved for a neutral atom
moving in a magnetic field.
PM93 analyzed various effects of the motion 
of the hydrogen atom 
in the perturbative regime, 
when $\kp$ is small.
When $\kp$ exceeds some critical value,
the magnetic and (motion-generated) electric fields 
form a potential well
at a distance $r_c=\mel\kp /\omega_c$ from the proton, which
gives rise to so-called {\em decentered} (or shifted)
states (Burkova et al.\ 1976). The decentered states
recover the cylindrical symmetry ($s$ becomes a good quantum number
again at large $\kp$), 
but the axis of symmetry, with respect to which $s$ is
defined, is shifted 
by $r_c$. In other words, these states can be described in the
{\it shifted\/} $s$-adiabatic approximation.
Binding energies, wavefunctions and oscillator
strengths of radiative transitions
between the discrete levels
of an arbitrarily moving H atom (beyond the
adiabatic approximation)
have been studied by Potekhin (1994; hereafter P94).
Our recent study of the hydrogen 
bound-bound opacity in NS atmospheres 
(Pavlov \& Potekhin 1995; hereafter PP95)
was based on these results. 

Less investigated are effects of the motion across the 
field on the bound-free radiative transitions of atoms. 
PM93 have shown, in the perturbation approach, 
that the motion strongly affects photoionization
cross sections through opening and enhancement of additional
photoionization channels forbidden for the non-moving atom due to
its cylindrical symmetry, and through the strong magnetic 
broadening of the photoionization edges.
Bezchastnov \& Potekhin (1994; hereafter BP94) 
suggested a modification of the adiabatic approach which allows for 
different displacements of the initial and final states and 
presented a convenient formalism for calculating
the transition rates between such states. 
They have also performed calculations with the initial state 
treated non-adiabatically; the two approaches exhibited 
a fair agreement with each other.  
Kopidakis, Ventura \& Herold (1996; hereafter KVH96) also went beyond the 
perturbation approximation
and calculated the frequency dependences of the cross sections
for various values of $\kp$, 
assuming these values small enough to neglect the 
decentered bound states.
The results of the two latter papers
differ from each other by orders
of magnitude
for radiation polarized {\it transversely\/} to the field. 
The origin of the discrepancy 
is the same as for atoms at rest --- KVH96 employed the velocity form
of the interaction potential
which led to wrong results when used with the $n$-adiabatic
approximation which treats the Landau number $n$ 
as the exact quantum number.
Both BP94 and KVH96 treated final (continuum)
states of electron in the (shifted) adiabatic approximation,
neglecting coupling of final states with different $s$ and 
contribution of quasi-bound (autoionizing) states to the
photoionization cross section.

The general non-adiabatic numerical approach developed 
by PPV97 for the continuum wave functions of the
strongly magnetized hydrogen
has been applied there to the particular 
case of non-moving atoms. In the present work, we employ that
approach, together with
the theoretical formalism of 
P94 and BP94, for an accurate 
treatment of both discrete and continuum states of the moving atom
and the radiative transitions between them. 
Unlike BP94 and KVH96, 
we do not restrict ourselves to the case of relatively small 
generalized momenta of atoms, 
but consider also the  decentered states 
which arise at large $\kp$. 
This enables us to calculate  
the cross sections of photoionization by arbitrarily 
polarized photons of the hydrogen 
atom arbitrarily moving in a magnetic field 
$B\sim 10^{12}-10^{13}$~G. Averaging these cross sections with
thermal distributions of 
atoms yields the bound-free
opacities needed for modeling NS atmospheres.

In Section 2 we outline basic techniques for numerical 
treatment of a strongly magnetized H atom and its interactions 
with radiation (the general form of the interaction matrix elements 
is given in 
the 
Appendix). In Section 3 we present the photoionization 
cross sections of H atoms in various states of motion across the 
magnetic field and 
cross sections averaged over thermal motion 
for conditions typical for atmospheres 
of isolated NSs.
\section{THEORETICAL FRAMEWORK}
\subsection{Generalized momentum and decentering of H atoms}
The center-of-mass motion of an atom in a magnetic field 
can be conveniently described by the generalized momentum $\vv{K}$
(Gorkov \& Dzyaloshinskii 1968; see also 
Johnson, Hirschfelder, \& Yang 1983),
\begin{equation}
\vv{K} = \vv{P} - \frac{e}{2c} \vv{B} \times (\vv{r}_e -\vv{r}_{\rm p}),
\end{equation}
where $ \vv{P}$ is the canonical momentum, 
$\vv{r}_{\rm e}-\vv{r}_{\rm p}$ is the relative
coordinate,
the subscript ``e'' or ``p'' indicates electron or proton, 
respectively,
and the cylindric gauge of the vector potential, 
$\vv{A} ( \vv{r} ) = \frac{1}{2} \vv{B} 
\times \vv{r} $, 
is used. 
We shall consider the representation in which 
all three components of $\vv{K}$ have definite values. Then the two-body 
wavefunction can be factorized into a phase factor 
depending on the center-of-mass coordinate and the wavefunction 
$\psi_{\vvsmall{K}}$ 
depending on $\vv{r}_{\rm e}-\vv{r}_{\rm p}$.

It is useful to define a basic electron-proton separation 
$\eta\vv{r}_c$ and to regard the deviation from it, 
$\vv{r} = \vv{r}_{\rm e} - \vv{r}_{\rm p} - \eta\vv{r}_c$, 
as an independent variable. Here
$\vv{r}_c = (c/eB^2) \vv{B}\times\vv{K}$ 
is the position of the relative guiding center, 
and $\eta$ is the {\it shift parameter\/}. 
It is also convenient to use a special axial gauge of the vector potential 
(Vincke, Le Dourneuf, \& Baye 1992; P94).  
Then we arrive at a set of Hamiltonians $H^{(\eta)}$ and 
wavefunctions $\psi^{(\eta)}(\vv{r})$ depending on 
$\eta$. In the particular 
cases of $\eta=0$ and $\eta=1$, the
``conventional'' and ``shifted'' 
representations of Gorkov \& Dzyaloshinskii (1968) are recovered. 
Wavefunctions with different $\eta$
are related to each other by the phase transformation (BP94)
\begin{equation}
   \psi^{(\eta)}_{\vvsmall{K}}(\vv{r}) = 
   \exp\left[-{{\rm i}\over\hbar}\,{\mpr-\mel\over 2\mH}\,
   \eta\vv{K}\vv{r}_\perp\right]\,
   \psi^{(0)}_{\vvsmall{K}}(\vv{r}+\eta\vv{r}_c), 
\label{transform}
\end{equation}
where $\vv{r}_\perp=(x,y)$ is the coordinate transverse to 
the field $\vv{B}$, 
and $\mH = \mel+\mpr$ is the total mass of the atom.

If there were no Coulomb attraction, then
the electron Landau number $n=0,1,2,\ldots$ and the $z$-projection of
the angular momentum of the relative motion $s \geq -n$
would be exact quantum numbers. In this case
the energy of the transverse
motion (with the
constant minimum energy subtracted) is
\begin{equation}
   E^\perp_{ns} = \left[\rule{0mm}{2.2ex} n+(\mel/\mpr)(n+s)\right]
\hbar\omega_c~,
\end{equation}
and the transverse part of the wave function 
can be described (for a given $\kpvec$)
by a Landau function
\begin{equation}
   \Phi_{ns}(\vv{r}'_\perp) = {1\over\am\sqrt{2\pi}}\,
   {\rm e}^{-{\rm i}s\phi}\,
   I_{n+s,n}\left({{r'}_\perp^2\over2\am^2}\right)~ 
\label{Landau}
\end{equation}
which depends on 
$\vv{r}'_\perp = 
 \vv{r}_{e\perp} - \vv{r}_{p\perp} - \vv{r}_c$
($=\vv{r}_\perp(\eta=1)$).
Here $\phi$ is the 
azimuthal angle of 
$\vv{r}'_\perp$,
$\am = (\hbar c/eB)^{1/2}$ is the magnetic length, 
$I_{nn'}$ is a Laguerre function (Sokolov \& Ternov 1968), 
and the $z$-projection $s$
of the angular momentum of the relative motion
is defined in the shifted reference frame, $\eta=1$
(since $\kpvec$ is definite, the electron and proton 
do not possess definite $z$-projections of the angular momenta 
separately from each other --- see Johnson et al.~1983).
Equation \req{Landau} implies
that after photoionization 
of a moving hydrogen atom,
when the photoelectron moves away to large distances
at which the Coulomb interaction can be neglected,
it acquires the displacement $\vv{r}_c$ from the proton in the
transverse plane.
\subsection{Atomic wavefunctions}
A wavefunction $\psi_\kappa^{(\eta)}$ of an atomic state 
$|\kappa\rangle$ can be expanded over the complete set of the 
Landau functions
\begin{equation}
   \psi_\kappa^{(\eta)}(\vv{r}) = 
   \sum_{ns} \Phi_{ns}(\vv{r}_\perp)\, g^{(\eta)}_{n,s;\kappa}(z)
\label{expan}
\end{equation}
(note that $\vv{r}_\perp$ depends on $\eta$). 
The adiabatic approximation would correspond to retaining
only one term in this expansion.

A {\em bound state} of the atom can be numbered as 
$|i\rangle = |n_i,s_i, \nu, \vv{K}\rangle$, 
where $n_i$ and $s_i$ relate to the leading term of the 
expansion \req{expan}, and $\nu$ enumerates longitudinal 
energy levels
\begin{equation}
   E^\|_i = E_i - E^\perp_{n_i s_i} 
\end{equation}
and controls the $z$-parity:
$g^{(\eta)}_{n,s;\kappa}(-z)=(-1)^\nu g^{(\eta)}_{n,s;\kappa}(z)$. 
For a moving atom, this way of 
numbering is unambiguous at $\gamma\ga 300$ (P94).

Since the transverse factors $\Phi_{ns}$ in equation \req{expan}
are known analytically, only the one-dimensional longitudinal
functions $g^{(\eta)}_{ns;\kappa}$ are to be found
numerically. For the bound states, numerical calculations
(Vincke et al.~1992; P94) yield
the following. At small transverse generalized momenta $K_\perp$,
the states $\nu=0$ remain tightly bound and {\it centered\/},
with the electron cloud concentrated around the proton.
For such states, the conventional representation ($\eta=0$)
is appropriate. At large $K_\perp$, the states are
{\it decentered\/}, with the electron
localized in
a potential well shifted apart from the proton. Then $\eta=1$ is
the apt choice. With growing $\kp$,
the transition from the centered tightly-bound states to the
decentered states occurs in a narrow range of $\kp$
near $K_{\rm c} \simeq \sqrt{2\mH|E^\||}$,
and the shifted ($\eta=1$) adiabatic approximation becomes fairly
good at $K_\perp\gg K_{\rm c}$.
For the hydrogen-like states $\nu\geq 1$, however,
the mean electron-proton separation grows steadily,
being close to $r_c$ at arbitrarily small $K_\perp$,
so that these states can be described by the shifted
adiabatic approximattion at any $K_\perp$.
Finally, at very large $K_\perp$
($\gg\gamma\hbar/\ab$, where $\ab$
is the Bohr
radius) the longitudinal functions
become oscillator-like for any states, corresponding to
a wide, shallow parabolic potential well
of a depth
 $\simeq e^2/r_c$ below the continuum boundary
(Burkova et al.\ 1976; P94).

A {\em continuum state} can be numbered as 
$|f\rangle = |n_f,s_f, E_f, I, \vv{K}'\rangle$, where 
$n_f$ and $s_f$ correspond to an open channel for the energy $E_f$, 
and $I=\pm1$ corresponds to a type of the final state --- e.~g., 
to a given $z$-parity or to a direction of electron escape, 
depending on asymptotic conditions. 
In the photoionization process 
the value of $\vv{K}'$  is constrained by the momentum conservation law: 
$\vv{K}' = \vv{K} + \hbar\vv{q}$, 
where $\vv{q}$ is a photon wavevector. 

For the continuum without motion ($K_\perp'=0$), 
a technique for calculating the longitudinal wavefunctions 
has been developed by PPV97.
It can be easily extended to 
the general case, $K_\perp'\neq 0$, by numerating 
the orbitals $ns$ in equation \req{expan} for the state 
$|f\rangle$ with single integers
$j$ such that larger values of $j$ correspond to higher 
energies $E^\perp_{ns}$.
Since the Coulomb interaction ceases to distort 
the transverse wavefunctions at infinity, 
the orbitals in equation \req{expan} 
become uncoupled at large $z$ provided that the full-shift 
representation ($\eta'=1$) of the continuum state is chosen. 
We shall adopt this choice hereafter. 

\subsection{Interaction with radiation}
The cross section for ionizing an atomic state $|i\rangle$ 
into a continuum state $|f\rangle$ 
by a photon with frequency $\omega$, wavevector $\vv{q}$, 
and polarization vector $\vv{e}$ is 
\begin{equation}
   \sigma_{i\to f} = \pi {e^2\over\hbar c}\,{{\rm Ry}\over\hbar\omega}\,
   \left({{\rm Ry}\over E_f^\|}\,{\mu\over\mel}\right)^{1/2}
   \!\! {L_z\ab}\,
   |\langle f |\hat{M}|i\rangle|^2, 
\label{crsct}
\end{equation}
where $\mu$ is the reduced mass,
$L_z$ is the $z$-extension of the periodicity 
volume of the final state, 
and $\hat{M}$ is the dimensionless interaction 
operator. 
For an arbitrarily moving atom of finite mass, the expression for
$\hat{M}$ is given in 
the Appendix. 
Using equation \req{expan} for both the initial 
and final atomic states, we obtain 
\begin{equation}
   \langle f|\hat{M}|i\rangle = \sum_{nsn's'} 
   \langle n's'\eta',f| \, \langle n's'\eta' 
   | \hat{M} | ns\eta \rangle_\perp \, 
   | ns\eta,i \rangle_\| ,
\label{m_expan}
\end{equation}
where $|ns\eta,\kappa\rangle_\|$ corresponds to 
the longitudinal wavefunction $g^{(\eta)}_{ns;\kappa}(z)$, 
and $| ns\eta \rangle_\perp$ corresponds to 
$\Phi_{ns}$. 
The interaction operator $\hat{M}$ 
in this equation depends on 
$\eta',\eta$, because the common phase factor 
(eq. [\ref{transform}]) has been included in it 
(see the Appendix). 

The Landau functions for the initial and final states 
in equation \req{m_expan} depend on differently displaced arguments
if $\eta'\neq\eta$. 
This is the case for the tightly bound initial states 
at $K_\perp < K_{\rm c}$, for which the apt choice of the shift 
parameter is $\eta=0$ (while $\eta'=1$). 
The different displacements
strongly complicate explicit integration in the inner (transverse) 
matrix elements. 

In order to reduce the initial and final wavefunctions 
in the same basis, KVH96 expressed the displaced Landau state as a series 
over the undisplaced states. This reexpansion 
involves an additional summation index which runs over a huge 
number of values if the relative displacement is large enough, which 
renders the reexpansion method impractical at high $\kp$.

The difficulty can be circumvented using the creation-annihilation 
operator formalism of BP94.
Analytical calculation of $\langle n's'\eta' 
   | \hat{M} | ns\eta \rangle_\perp $
is performed in 
the 
Appendix. 
As a result, the interaction matrix element 
$\langle f|\hat{M}|i\rangle$ 
reduces to a sum of one-dimensional quadratures (eq.\,[\ref{m_final}]) 
feasible for numerical evaluation. 

\section{RESULTS AND DISCUSSION}
\subsection{Photoionization at fixed ${\bf K}_\perp$}    

\begin{figure}
\plotone{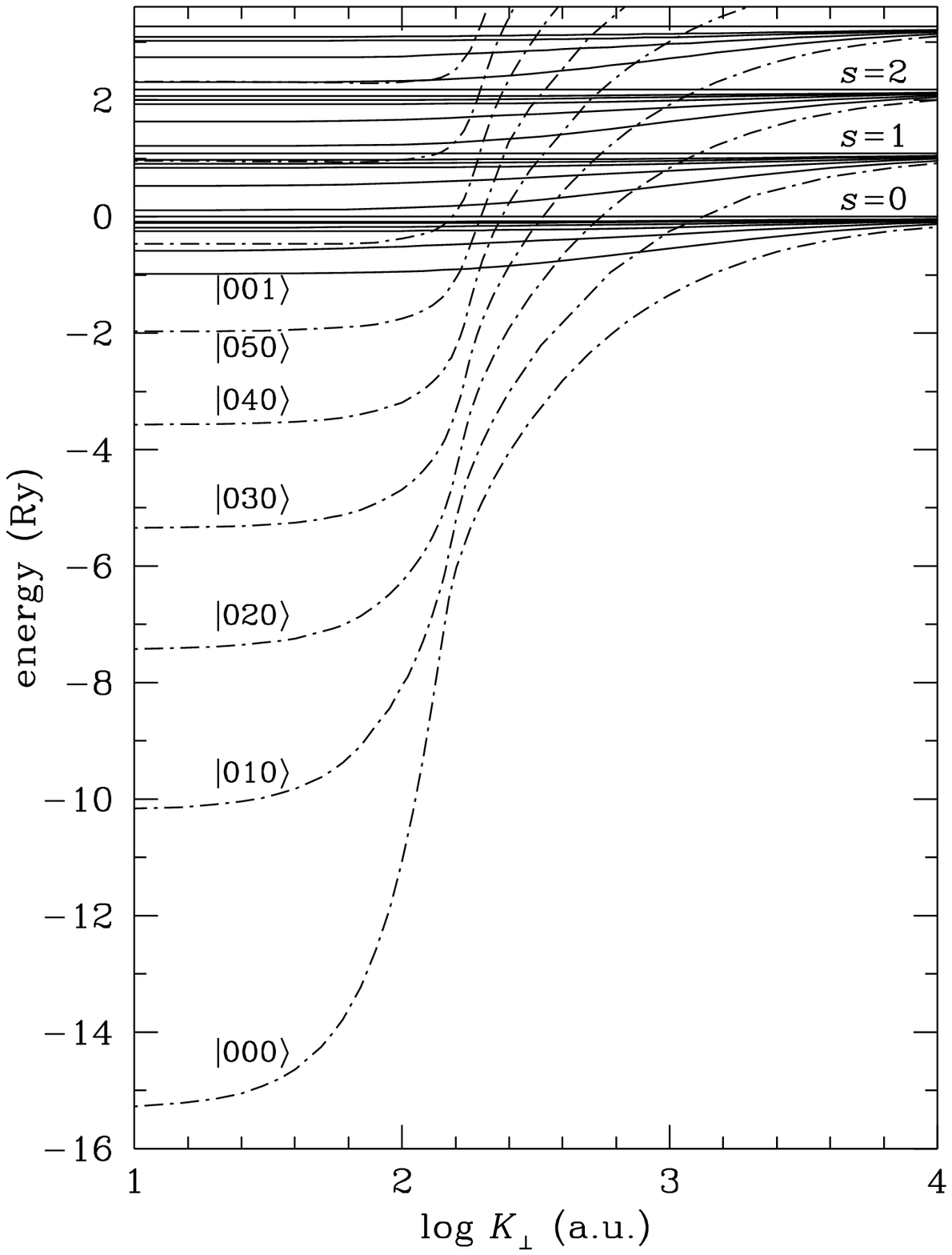}
\caption{
Energy spectrum ($E_{0s\nu}-E^\perp_{00}$)
vs.\ transverse generalized momentum $K_\perp$
at $B=2.35\times10^{12}$~G.
The curves
(dot-dashed for tightly-bound states and solid for hydrogen-like
states) 
 are labeled with
the quantum numbers, $|0s\nu\rangle$.
\label{fig1}}
\end{figure}

Many properties of the photoionization cross section
can be understood from the $\kp$-dependence of 
the level energies shown in Figure \ref{fig1}. 
The monotonic 
increase of the energies
shifts the photoionization thresholds redward. 
Admixture of many orbitals near ``anticrossings''
of levels
belonging to different $s$-manifolds
(e.~g., at $K_\perp \approx 150$ a.u.) causes peculiarities 
in the radiative transition rates (P94). 
The levels which lie above zero in Figure \ref{fig1} correspond to 
quasi-bound states in the continuum. 
They give rise to sharp resonances in the cross sections 
at the photon energies corresponding to 
transitions 
between the initial bound state and the quasi-bound states.
 
\placefigure{fig1}

\begin{figure}
\plotone{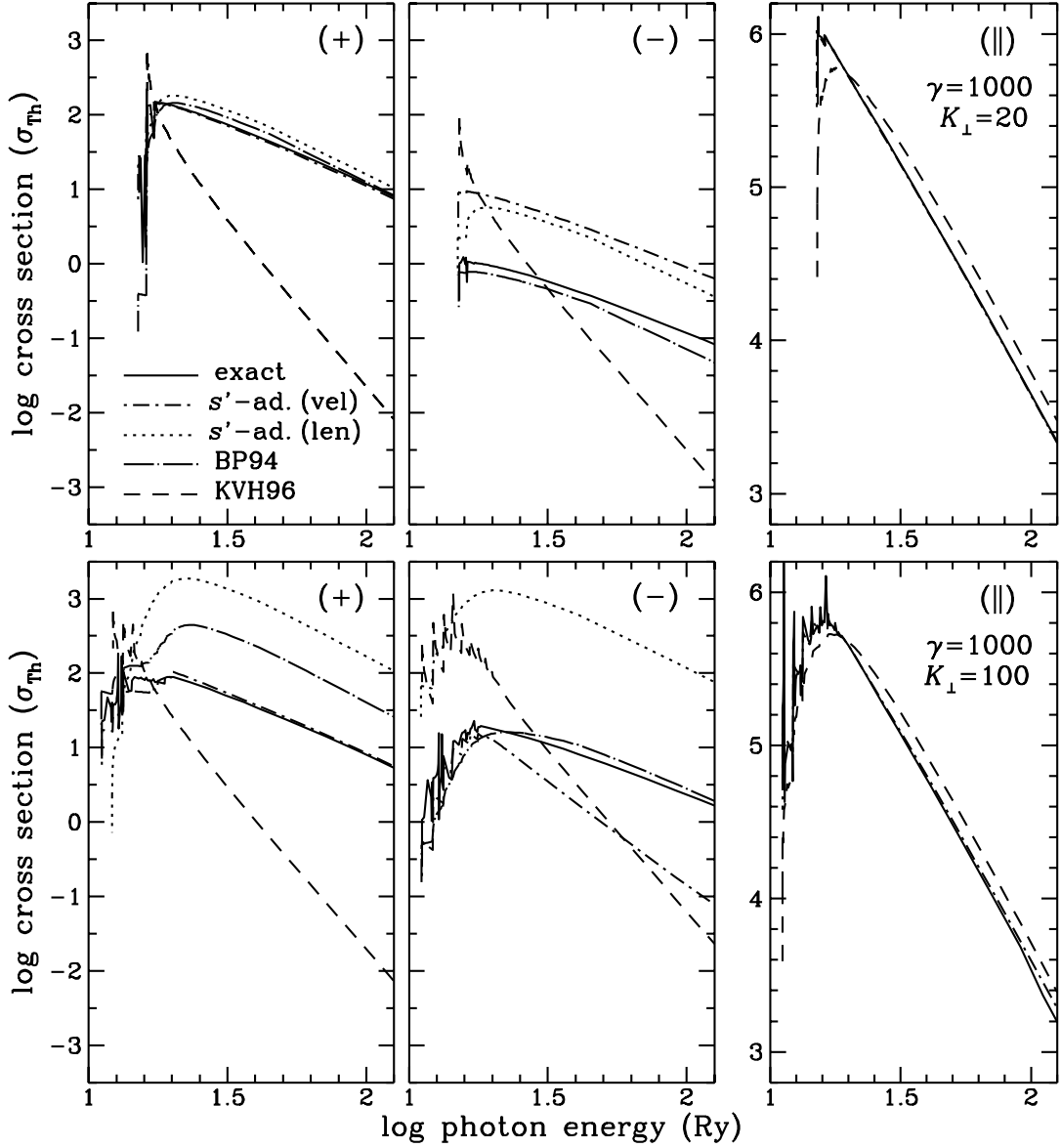}
\caption{
Energy dependence of the photoionization cross sections
(in units of the Thomson cross section $\sigma_{\rm Th}$)
of the ground-state H atom moving across the magnetic
field $B=2.35\times10^{12}$~G with the generalized momentum
$\kp=20$~a.u. ($\kappa\equiv\kp\am/
(\hbar\protect\sqrt{2})=0.447$, upper panels) 
and $\kp=100$~a.u. ($\kappa=2.236$, lower panels)
for the right ($+$), left ($-$), and longitudinal ($\|$)
polarizations (the incident photons move along the magnetic
field in the former two cases, and across the field in the latter
case).
Our numerical results (solid lines)
are compared with several common approximations (see text).
\label{fig2}
}
\end{figure}

Figure~\ref{fig2} demonstrates a comparison of 
our numerical results (solid lines) with those obtained 
using approximations encountered in literature. 
The short-dash-dot curves are obtained 
using the velocity form of the interaction operator
with the $s'$-coupling  ``switched-off''
($s'$-adiabatic approximation). 
We see that this coupling is very important 
for evaluating the cross section $\sigma_-$
(for the left circular polarization transverse to
the magnetic field) and less important 
for the other two basic cross sections, 
$\sigma_+$ (right circular polarization transverse to the field) and
$\sigma_\|$ (linear polarization along the 
field). 
The dotted lines are obtained using
the shifted adiabatic approximation for the final state 
and the length representation of the interaction operator.
Note that they are close to the short-dash-dot lines
in the upper panels.
This is in accordance with approximate equivalence of 
using the $n$-adiabatic approximation ($n,n'=0$)
with the length form and
using the velocity form with allowance 
for $n,n'>0$. This equivalence has been suggested by PP93 
and proved by PPV97 under the condition that the non-adiabatic 
corrections were small. It fails for $\kp\sim K_{\rm c}$
if the $s'$-coupling is neglected (compare dotted and
dot-short-dash curves in the lower panels),
but we have checked numerically that it holds if the $s,s'$-coupling 
is included.

The dot-long-dash curves correspond to the approach of BP94.
It differs from the plain shifted $(n's')$-adiabatic approximation
(dotted curves) by employing the orthogonalization 
of the final state with respect to the initial one 
(see eq.~[32] of BP94). 
At relatively small $\kp$ (upper panels), this approximation 
yields qualitatively correct results.
With increasing $\kp$, the approximate
$\sigma_+$ becomes progressively less
accurate due to violation of the cylindrical symmetry
of the continuum states of the atom; 
for instance, it exceeds our cross section 
by a factor of $\sim 5-7$
at $\kp =100$ a.u.
However, the approximate 
$\sigma_-$ and $\sigma_\|$
remain surprisingly accurate.

\placefigure{fig2}

Short dashes correspond to the approach of KVH96,
who use the $n$-adiabatic approximation 
for the initial state, and the shifted adiabatic and 
Born approximations for the final state. 
If to abandon the Born approximation 
(retaining the adiabatic one), then
$\sigma_\pm$ would become considerably lower,
by 0.5 to 3 orders of
magnitude for the parameters chosen in Fig.~\ref{fig2}, 
and even more different from the true cross sections.
This lowering is due to the approximate orthogonality 
of the initial and final {\it longitudinal\/} wave functions,
which is not provided by the Born approximation. 

All the discussed approximations miss the Beutler--Fano resonances 
(narrow peaks and dips near the thresholds), which occur 
at the energies of the quasi-bound states in the continuum. 
These resonances are analogous to 
those discussed by PPV97, 
except that now these quasi-bound states
all belong to the ($n'=0$)-manifold. 
Autoionization in this case becomes possible 
due to the motion-induced coupling of different $s'$-channels. 

Except for the missing resonances, the discussed approximations 
appear to be sufficiently accurate for the longitudinal ($\sigma_\|$) 
polarization. 

\begin{figure}
\plotone{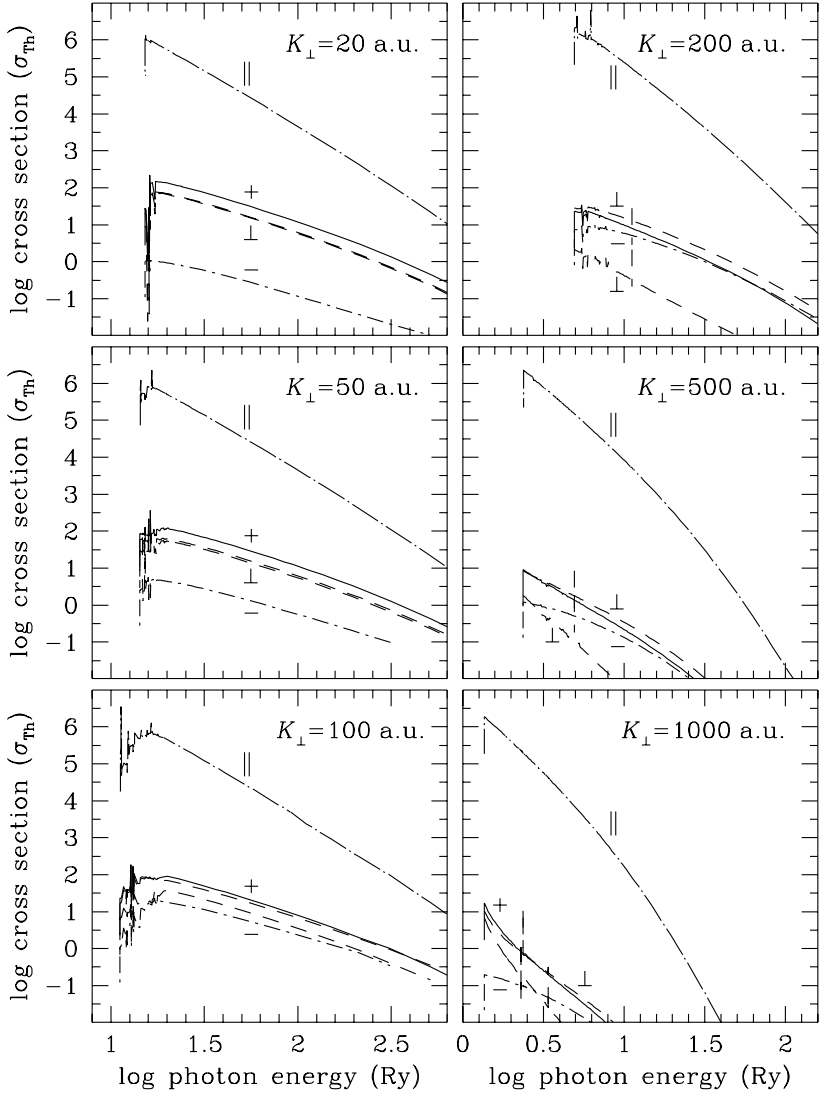}
\caption{
Photoionization cross sections 
of the ground-state H atoms moving with different
$K_\perp$ in the magnetic field $B=2.35\times10^{12}$~G.
Solid and dash-dot lines correspond to the right $(+)$
and left $(-)$ circular polarizations of photons propagating
along the magnetic field; long-dash-dot and dashed
lines correspond to the linear polarizations parallel ($\|$)
and perpendicular ($\perp$) to the field, for the transverse
propagation.
Upper and lower dashed curves are for the wavevector parallel
and perpendicular to the transverse component of the generalized
momentum.
\label{fig3}
}
\end{figure}

Figure~\ref{fig3} demonstrates
a representative sample 
of our numerical results 
for $\gamma = 1000$. 
In accordance with Figure 1, the photoionization thresholds
shift redward with increasing $\kp$.
Since  $K_{\rm c}\approx 150$ a.u. 
at this field strength,
the left and right panels correspond to the centered 
and decentered atoms, respectively. 

\placefigure{fig3}

For the centered atoms, the slopes of the curves
for each given polarization 
remain practically coincident with 
those at $\kp=0$. The values of the cross sections for the parallel 
and right polarizations, corresponding to 
the radiative transitions 
with 
$s'=s$ and $s'=s+1$ allowed at $\kp=0$ 
(when $s$ is a ``good'' quantum number), 
decrease only slightly with increasing $\kp$ from 0 to $K_{\rm c}$. 
For the left polarization, 
radiative transitions from the ground state 
are forbidden ($\sigma_-=0$) at $\kp=0$ (PP93). 
With growing $\kp$, the coupling between different 
$s$-channels arises, and $\sigma_-$ grows 
(approximately as $\kp^2$ --- see PM93)
until it becomes 
comparable to $\sigma_+$ at $\kp\sim K_{\rm c}$. 

For the {\it decentered\/} atoms (right panels of Fig.~\ref{fig3}), 
the energy dependences
steepen with growing $\kp$. This is explained by the change of  
the effective potential from Coulomb-like at $\kp\lesssim K_{\rm c}$ 
to oscillator-like as the atom becomes decentered. 
The effective potentials corresponding 
to different $s$ become nearly identical, that results in a significant 
decrease of $\sigma_+$. Still faster is the decrease 
of $\sigma_-$ caused 
by restoring applicability of the $s$-adiabatic 
approximation and corresponding selection rules 
at large $\kp$. 
At the same time, the threshold value of $\sigma_\|$ remains 
almost independent of $\kp$. 

\begin{figure}
\plotone{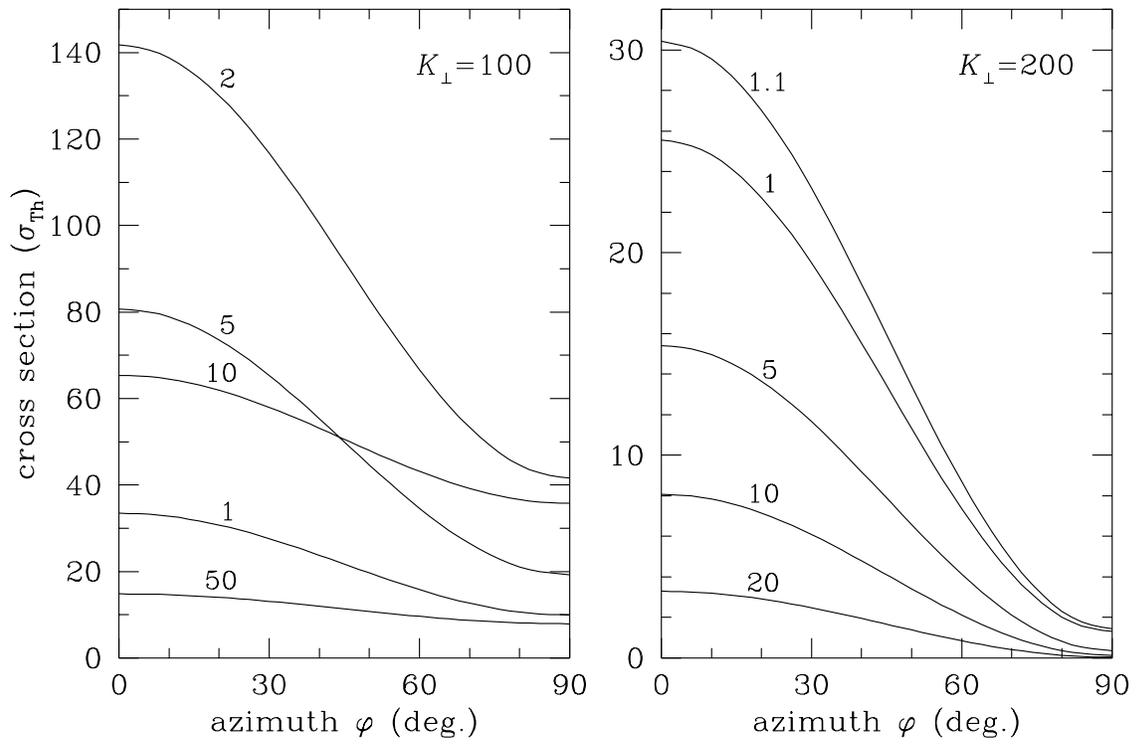}
\caption{
Angular dependence of the photoionization cross section
$\sigma_\perp$
of the ground-state H atoms moving with $K_\perp=100$ and 200 a.u.
in the magnetic field $B=2.35\times10^{12}$~G.
The wavevector and polarization direction are
perpendicular to the magnetic field.
Numbers near the curves indicate energies
$E_f$ of the final state (in Ry).
\label{fig4}
}
\end{figure}

Since the moving atom loses its cylindrical symmetry,
the cross sections depend on the 
angle $\varphi$ 
between $\kpvec$ 
and the transverse component of the wavevector. 
The $\varphi$-dependence is most pronounced 
when both the wavevector $\vv{q}$ and polarization direction
$\vv{e}$ are
perpendicular to $\vv{B}$. 
The cross sections $\sigma_\|$ and $\sigma_\perp$ 
are drawn in Figure \ref{fig3} by dot-long-dash and dashed lines, 
respectively. The former cross section does not depend on $\varphi$
in the dipole approximation, whereas the small 
non-dipole corrections (eq.~[\ref{mz}]) 
cause a weak dependence with an amplitude $\ll1$\%. 
Therefore, we show only one curve for $\sigma_\|$ 
on each panel.
For the transverse polarization, 
the $\varphi$-dependence becomes 
pronounced at 
$\kp\gtrsim\hbar/\am\approx30$ a.u. 
The upper and lower dashed curves correspond to the largest
and smallest possible 
values of $\sigma_\perp(\varphi)$,
at $\varphi=0$
and $\varphi=90\arcdeg$, respectively. 
The reason is that the transversely polarized photons 
``see'' the largest transverse
size of the atom (stretched perpendicular to $\kpvec$)
 when they propagate along $\kpvec$, 
and the smallest one when propagating transversely to $\kpvec$. 
The $\varphi$-dependence between these two extremes, 
illustrated in Figure \ref{fig4}, 
is described by the formula
\begin{equation}
   \sigma_\perp(\varphi)=\frac{\sigma_+ + \sigma_-}{2} +
   A \cos2\varphi~.
\label{phi}
\end{equation}
Note that in the absence of coupling of different $s$-channels 
the amplitude $A$ in equation \req{phi} would equal 
$\sqrt{\sigma_+\sigma_-}$, which is always greater than the 
actual value of $A$, calculated numerically. 

\placefigure{fig4}

\begin{figure}
\plotone{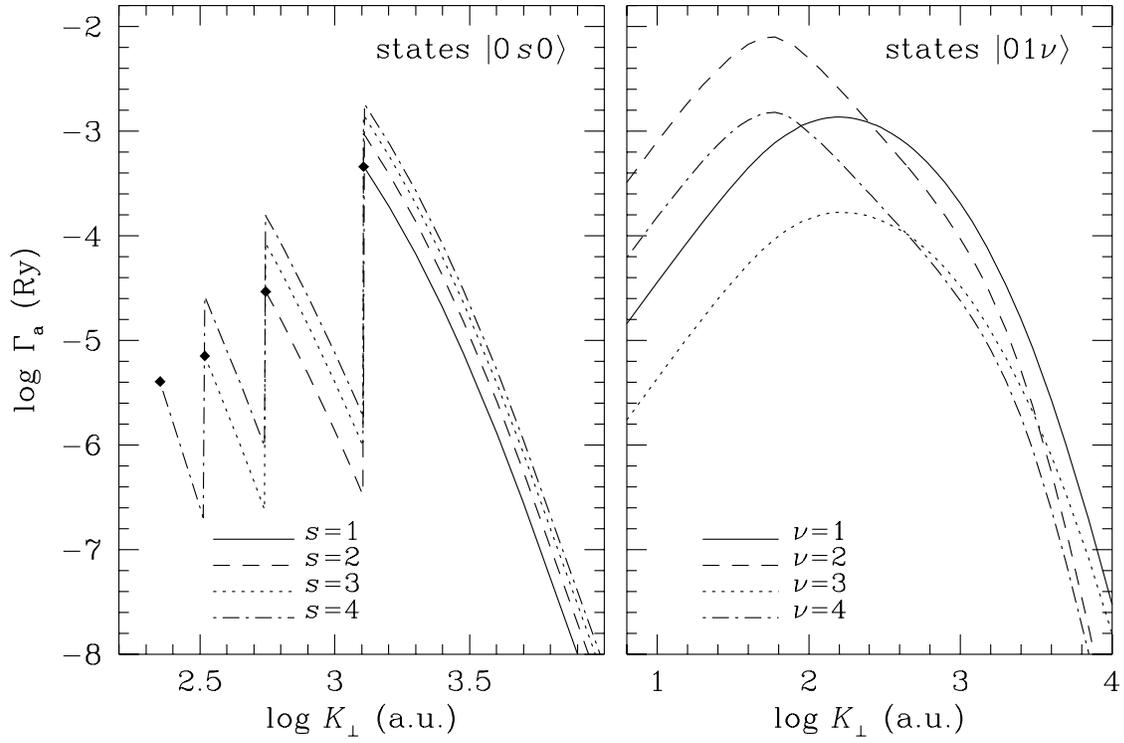}
\caption{
Autoionization widths of several tightly-bound
(left panel) and hydrogen-like (right panel)
atomic resonances as function of $\kp$
for $B=2.35\times 10^{12}$ G.
Each curve on the left panel starts
from the point where the coresponding state
enters continuum.
\label{fig5}
}
\end{figure}

The curve corresponding to $E_f=1.1$ Ry in Figure \ref{fig4} 
goes higher than that corresponding to $E_f=1$ Ry because the
ionization channel $s'=1$ opens at 
$E_f=\hbar\omega_{c\rm p}=1.089$ Ry,
where $\omega_{c\rm p}=(\mel/\mpr)\omega_c$ is the 
proton cyclotron frequency.
The curve corresponding to $E_f=2$ Ry goes 
much higher than the other curves because it falls 
onto a slope of a Beutler--Fano resonance. 
Such resonances are most prominent at $\kp\sim K_{\rm c}$ 
(cf.\ Fig.~\ref{fig3})
because of the violation of the $s$-adiabatic approximation 
in this range of the transverse generalized momenta. 
Nevertheless, these peaks remain very narrow. 
Figure~\ref{fig5} presents 
some autoionization widths $\Gamma_{\rm a}$ 
calculated according to 
equations (14) and (15) of PPV97
(with obvious generalization to the case of moving atoms).
The left panel shows the widths of the tightly-bound 
states $s=1$ through 4 which enter the continuum 
at $\kp\approx 1200$, 560, 340, and 237 a.u., respectively 
(see Fig.~\ref{fig1}). Owing to the similarity of 
the effective potentials related to different $s$ values, 
the jumps of the curves corresponding to opening 
of additional autoionization channels occur at 
nearly the same values of $\kp$. 
The hydrogen-like levels of the ($s=1$)-manifold 
(right panel of Fig.~\ref{fig5}) belong to the continuum 
at any $\kp$. The autoionization widths 
vanish at $\kp\ll K_{\rm c}$ and at $\kp\gg K_{\rm c}$ 
because of restoring cylindrical symmetry of the wavefunctions. 

\placefigure{fig5}

\subsection{Absorption by atoms in thermal equilibrium}
The bound-free absorption coefficient for a hydrogen gas of given
temperature and density is proportional to the photoionization
cross section $\sigma_{\kappa\to\kappa'}(\vv{K}_\perp )$ folded with 
$N_\kappa(\vv{K})$,
the distribution of the number density of atoms in the initial state.
Since the level energy $E_\kappa(\kp)$ tends to
a constant value at $\kp\rightarrow\infty$,
the normalization integral for the Boltzmann distribution
$N_\kappa(\vv{K})\propto \exp[-E_\kappa(\kp)/k_{\rm B}T]$
diverges.
The divergence is eliminated if nonideality of the gas is taken
into account.
At high $\kp$, the binding energies become small,
and atomic sizes become large. This
means that the atom can be easily destroyed by surrounding particles,
so that very high $\kp$ should not contribute to the corresponding
integrals. In order to allow for this effect, we use
the occupation probability formalism, described in detail
by Hummer \& Mihalas (1988) and generalized to the case of the
$\kp$-dependent atomic structure by PP95.
Following this approach, we write
\begin{equation}
   N_\kappa(\vv{K})\propto
   w_\kappa(\kp)\exp [-E_\kappa(\kp)/k_{\rm B}T],
\end{equation}
   where $w_\kappa(\kp)$ is the occupation probability, 
   equal to the fraction of atoms 
   (in the specified state $|\kappa,\vv{K}\rangle$) which 
   essentially preserve quantum-mechanical 
   (particularly, optical) properties of an isolated atom. 
We estimate $w_\kappa(\kp)$ according to equation (14) of PP95,
based on the quantum-mechanical atomic sizes calculated by P94.
A rapid decrease of $w_\kappa(\kp )$ at $\kp \ga \gamma\hbar/\ab$
(see Fig.~3 of PP95) limits contribution from atoms with
large generalized momenta and provides convergence of integrals
over $\kp$. 

Since the threshold energies decrease with increasing $\kp$,
folding the photoionization cross sections with  $N_\kappa (\kp )$
leads to the redward broadening of the photoionization edges,
with a typical width $\Gamma_M\sim k_BT$ (PM93).
It has been shown by PM93 and PP95 that the ``magnetic
width'' is much greater than the Doppler and (impact)
collisional widths
under typical conditions of NS atmospheres. In this paper 
we take full account of the magnetic broadening 
and neglect other broadening mechanisms. 

\begin{figure}
\plotone{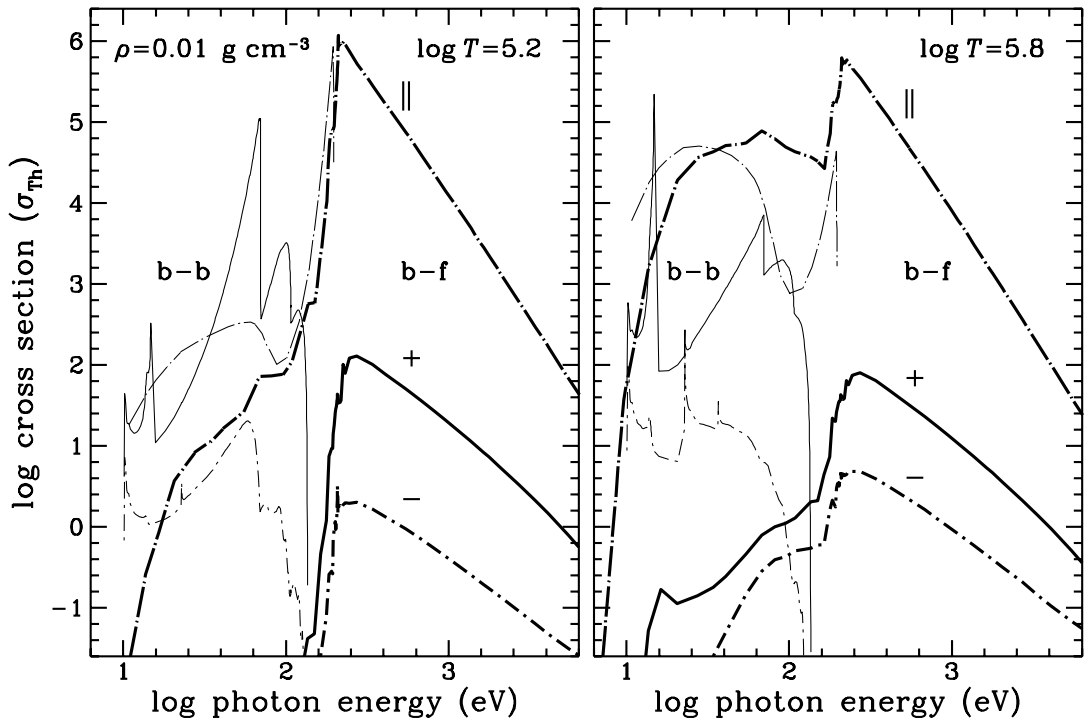}
\caption{
Photoionization cross sections (heavy lines)
for three polarizations (drawn with the same line styles
as in Fig.~\protect{\ref{fig3}}),
averaged over the thermal
distribution of atoms at the magnetic field $B=2.35\times 10^{12}$ G,
density $\rho=0.01$ g cm$^{-3}$, and temperatures
$T=10^{5.2}$~K and $10^{5.8}$~K. Light lines
represent cross sections arising from the bound-bound transitions (PP95).
\label{fig6}
}
\end{figure}

The averaged photoionization cross sections 
for atoms on the ground level
are shown in Fig.~\ref{fig6} for two values of temperature $T$. 
For comparison, we show also 
cross sections 
for the bound-bound (b-b) transitions from the ground level (PP95). 
Both the b-b and bound-free (b-f) averaged cross sections are
corrected for the stimulated emission, i.~e.,
multiplied by $(1-\exp[-\hbar\omega/k_BT])$.
For the circular polarizations, the b-b and b-f transitions 
dominate in different (adjacent) spectral ranges ---
the low-energy b-f component, arising from 
the decentered states (cf.\ Fig.~\ref{fig3}),
turns out to be much weaker 
than the b-b component in the same energy range. 
For $\sigma_\|$, this is also the case at $\log T=5.2$,
but not at $\log T=5.8$. In the latter case, the low-energy 
b-f component dominates the absorption spectrum, 
except at very low energies $\hbar\omega\lesssim 40$ eV. 

\placefigure{fig6}

The peak of $\sigma_+$ 
in the right panel of Figure \ref{fig6} 
at $\hbar\omega\approx 15$ eV 
and the corresponding peaks of the `b-b' curves 
arise near the proton cyclotron frequency
$\omega_{c\rm p}$,
around which 
the transitions of strongly decentered atoms 
are concentrated (cf.\ PP95). 

\begin{figure}
\plotone{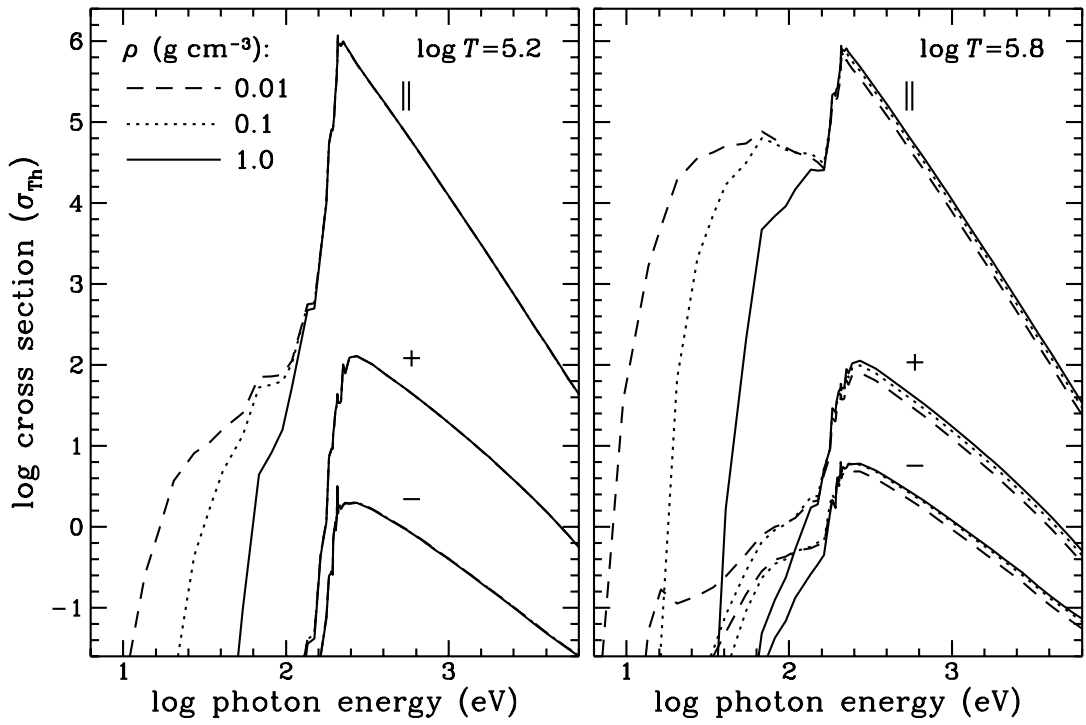}
\caption{
Averaged photoionization cross sections
for three polarizations,
at $B=2.35\times 10^{12}$ G, three values of $\rho$,
and two values of $T$.
\label{fig7}
}
\end{figure}

Figure~\ref{fig7} demonstrates the effect of density on 
the averaged photoionization cross sections. With increasing $\rho$, 
the low-energy component 
rapidly decreases because the decentered atoms, 
due to their large sizes, are 
easier destroyed by the plasma microfields. 
The growth of the high-energy component
noticeable at higher $T$ is explained by the
relative increase of the number of centered atoms
when the decentered atoms are destroyed.
It should be noted that at high densities one
more effect, broadening of the photoionization edges
by quasi-static interactions with surrounding particles, may become important.
As the density grows,
highly excited bound states become randomly disturbed
by the surrounding and
form a quasi-continuum. This results in continuous photon
absorption below the undisturbed photoionization threshold
(e.~g., D\"appen, Anderson, \& Mihalas 1987).
These ``bound-quasi-free'' transitions
require a separate treatment and are not included in 
the present paper.

\placefigure{fig7}

\begin{figure}
\plotone{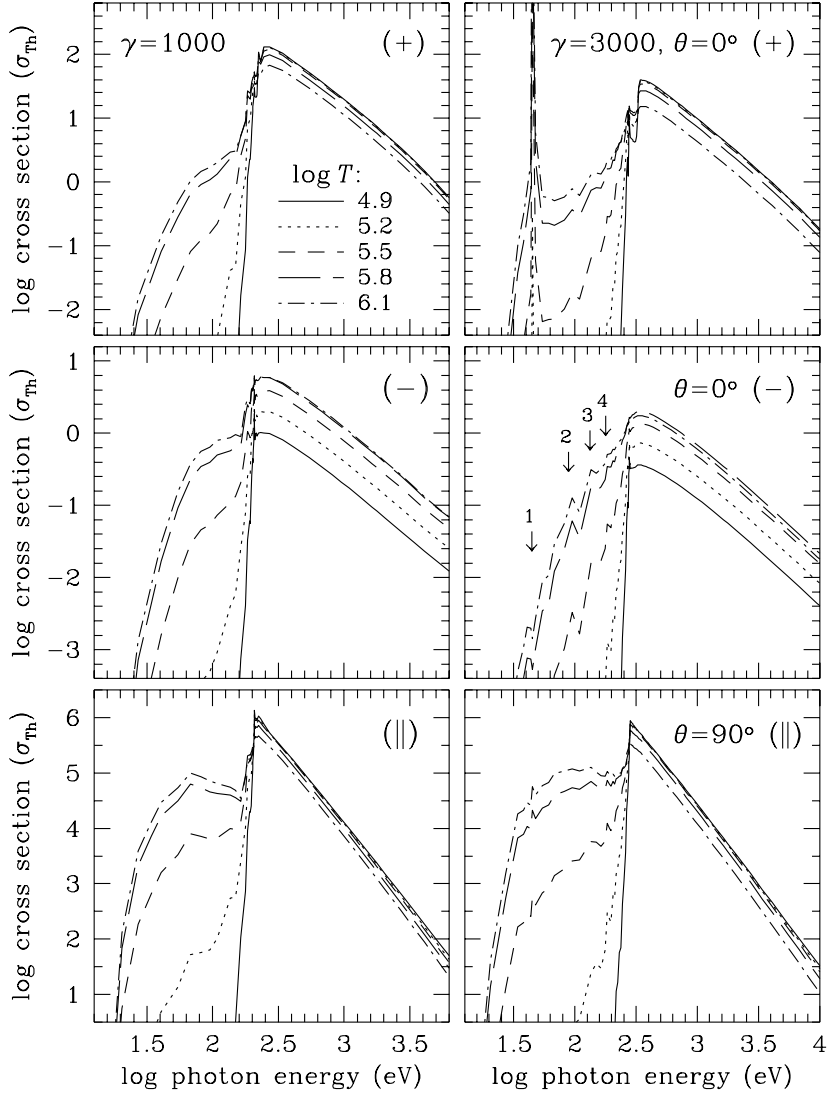}
\caption{
Averaged photoionization cross sections
for the right (top panels), left (middle panels),
and longitudinal (bottom panels) polarizations
for $\rho=0.1$ g cm$^{-3}$,
$B=2.35\times 10^{12}$~G and $7\times 10^{12}$~G
(left and right panels, respectively), and
different temperatures.
\label{fig8}
}
\end{figure}

Figure~\ref{fig8} shows the effects of $T$ and $B$ on 
the photoabsorption spectral shapes. 
With growing $T$, the low-energy components significantly 
increase, and the high-energy components of $\sigma_\|$ and
$\sigma_+$ decrease slightly, because the decentered states
become more populated.
At lower $T$,
the high-energy component of $\sigma_-$ grows with $T$
linearly,
\begin{equation}
\sigma_-\approx \frac{\hbar\omega_{c\rm p}\, k_BT}{
2\epsilon_{000,010}^2}\,\,
\sigma_+~,
\end{equation}
in accordance with the result of PM93 obtained in the perturbation
regime
($\epsilon_{000,010}$ is the energy of the main b-b transition
for the right polarization -- e.~g., $\epsilon_{000,010}\simeq 70$ eV
for $\gamma=1000$). At higher $T$, the growth of $\sigma_-$ decelerates
because of decreasing fraction of the centered states and
nonperturbative effects.

At higher $B$, the above-discussed proton cyclotron resonance 
for $\sigma_+$ becomes stronger. Owing to 
the approximate similarity of the energy spectra 
related to different $s$-manifolds (see Fig.~\ref{fig1}), 
the individual Beutler--Fano resonances, 
although smoothed by the thermal averaging, 
are collected together into regularly 
placed peaks. Such peaks corresponding 
to four proton cyclotron harmonics 
are indicated by arrows for $\sigma_-$ at $\gamma=3000$ 
in Fig.~\ref{fig8}. At high densities, the peaks can be
further smoothed by interaction with surrounding particles.

\placefigure{fig8}

\section{CONCLUSIONS}
The results presented here provide, together with the
results of PP95 on the bound-bound transitions, a basis for calculations
of realistic opacities of NS atmospheres. To find
the opacities, one should add transitions from excited atomic
levels, which is straightforward. More difficult are the
problems of ionization equilibrium in the dense, non-ideal
plasmas and closely related problem of the non-ideality
effects on the bound-free transitions (including transitions
to the quasi-free states below the unperturbed photoionization threshold).
Finally, polarizations of the normal modes
for any frequency and direction of propagation should be found
for solving the radiative transfer equations. 
Bulik \& Pavlov (1996) have described how these polarizations
can be calculated if the basic cross sections $\sigma_\pm$
and $\sigma_\|$ are known.

Our results show that atomic motion in strong magnetic
fields drastically changes
both the bound-bound and bound-free opacities. In particular,
it  considerably broadens the spectral lines
and photoionization edges (with a typical magnetic width
$\sim k_BT$), gives rise to additional absorption at low
frequencies due to decentered atoms, and opens new transition
channels forbidden for atoms at rest. These unusual properties
of the spectral opacities  alter the structure of the
NS atmospheres and manifest themselves in the
spectra and light curves of radiation emergent from the 
NS surface layers. 

\acknowledgments
The authors are grateful to
Victor Bezchastnov and Joseph Ventura for fruitful
discussions, and to Cole Miller and the referee, Tomek Bulik, for 
useful remarks. 
This work was partially supported through NASA grant
NAG5-2807, INTAS grant 94-3834, RBRF grant 96-02-16870a,
and DFG--RBRF grant 96-02-00177G.
AYP gratefully acknowledges the hospitality of 
Joseph Ventura at the University of Crete, 
Gilles Chabrier at ENS-Lyon, 
and Chris Pethick at Nordita,
where a part of this work has been done. 

{ 
\appendix
\onecolumn
\addtocounter{section}{1}
\section*{APPENDIX: INTERACTION MATRIX ELEMENTS}
Velocity and length representations of the operator 
of interaction of a moving hydrogen atom with radiation 
have been derived by BP94.
Here we employ the velocity representation as the simpler one. 
KVH96 have noted that 
the interaction of the radiation field with the electron spin 
should be taken into account. It has been found by PPV97  
that the spin term almost does not change the 
cross sections 
at energies below 10 keV. Nevertheless, we include this term 
for the sake of generality,
neglecting the terms responsible for the spin-flip transitions 
(Schmitt et al.\ 1981) which are strictly 
forbidden at the considered photon energies below $\hbar\omega_c$. 

In the conventional representation of the wavefunctions, 
equation (A6) of BP94 yields
for the dimensionless interaction 
operator $\hat{M}$:
\begin{equation}
   (2\ab)^{-1} \hat{M} = \exp\left({\rm i}
   {\mpr\over\mH}\vv{q}\vv{r} \right)
   \left[ \left({\vv{\pi}\over\hbar} 
   + {\mel\over\mH}{\vv{K}\over\hbar} 
   + {\vv{q}\over 2} \right)
   \vv{e}
   -
   {{\rm i}\over 2}(\vv{q}\times\vv{e})_z \right] 
   +
   \left({\mel\over\mpr}{\vv{\Pi}\over\hbar} 
   - {\mel\over\mH}{\vv{K}\over\hbar} 
    \right)
   \vv{e},
\end{equation}
where
$\vv{q}$ is the photon wavevector, $\vv{e}$ is its 
polarization vector, and
negligibly small terms $\sim(\mel/\mpr)q$ are omitted.
Here $\vv{r} = \vv{r}_{\rm e} - \vv{r}_{\rm p}$ 
is the relative coordinate, and $\vv{\pi}$ and $\vv{\Pi}$ 
are the electron- and proton-type kinetic momentum operators: 
\begin{mathletters}
\begin{eqnarray}
   \vv{\pi} = \vv{p} + {e\over 2c} \vv{B}\times\vv{r}, 
\label{pi}\\
   \vv{\Pi} = \vv{p} - {e\over 2c} \vv{B}\times\vv{r}.
\end{eqnarray}
\end{mathletters}
Their cyclic components act on the Landau states as 
\begin{mathletters}
\begin{eqnarray}
   \pi_\pm |ns\rangle_\perp &=& \mp{{\rm i}\hbar\over\am}\,
   \sqrt{n_\ast}\,| n\pm1,s\mp1 \rangle_\perp,
\label{pi_pm}\\
   \Pi_\pm |ns\rangle_\perp &=& \mp{{\rm i}\hbar\over\am}\,
   \sqrt{n_\ast+s}\,| n,s\mp1 \rangle_\perp,
\label{Pi_pm}
\end{eqnarray}
\end{mathletters}
where $n_\ast = n+1$ and $n_\ast =n$ for the right (+) and 
left ($-$) 
components, respectively. 
If the full-shift representation, $\eta'=1$, is used for the final state, 
and an arbitrary shift parameter $\eta$ for the initial one, 
then, according to equation (A15) of BP94, 
\begin{eqnarray}
   \hat{M} &=& {2\hbar\over e^2} 
   \exp\left[-{{\rm i}\over 2}\,{\mpr-\mel\over\mH}
   (1-\eta)\vv{K}\vv{r}_\perp \right]
\nonumber\\
   &\times& 
   \vv{e}\left(
   \exp\left[\rule{0mm}{2ex}{\rm i}(\vv{q}\vv{r}_\perp/2 + q_z z) 
   \right] 
   \vv{F}_{\rm e} 
   +
   \exp\left[
   -{\rm i}\eta\vv{r}_c \vv{q} - {\rm i}\vv{q}\vv{r}_\perp/2
   \rule{0mm}{2ex}
   \right]
   \vv{F}_{\rm p}
   \right),
\label{m_shift}
\end{eqnarray}
where $\vv{r} = \vv{r}_{\rm e} - \vv{r}_{\rm p} - \eta\vv{r}_c$. 
Once again, the terms $\sim q \mel/\mpr$ are 
omitted.
The components of the vector operators $\vv{F_{\rm e}}$
and $\vv{F}_{\rm p}$ are
\begin{mathletters}
\begin{eqnarray}
   F_{{\rm e}+} = {\pi_+\over\mel} + (1-\eta){K_+\over\mH} + 
   {\hbar q_+\over\mel},
& ~ &
   F_{{\rm e}-} = {\pi_-\over\mel} + (1-\eta){K_-\over\mH}, 
\\
   F_{{\rm p}\pm} = {\Pi_\pm\over\mpr} - (1-\eta){K_\pm\over\mH}, 
&&
\\
   F_{{\rm e}z} = {1\over\mel}(p_z + \hbar q_z/2),
&&
   F_{{\rm p}z} = p_z/\mpr .
\end{eqnarray}
\end{mathletters}
In order to proceed with equation \req{m_expan}, we need to calculate 
the transverse matrix elements
$\langle n's'\eta' | \hat{M} | ns\eta \rangle_\perp $. 
In the particular case of the dipole approximation ($q\to 0$), 
such calculation has been performed by BP94. 
In the general case, we use the following result obtained 
with the same operator technique as in BP94:
\begin{equation}
   \langle n's'\eta',\kpvec+\vv{q}_\perp | \exp\left[-{{\rm i}\over 2}\,
   {\mpr-\mel\over\mH}\,(\eta'-\eta)\vv{K}\vv{r}_\perp 
   + {\rm i}\tilde{\vv{q}}\vv{r}_\perp/2 \right]
   | ns\eta,\kpvec \rangle_\perp 
   =
   J_{nn'}(\xi_{\rm e}) J_{n+s,n'+s'}(\xi_{\rm p}),
\label{jj}
\end{equation}
where $\tilde{\vv{q}}$ is arbitrary, 
\begin{mathletters}
\begin{eqnarray}
   \xi_{\rm e} &=& -{\rm i}\am \left(
   {\mel\over\mH} (\eta'-\eta)K_+/\hbar + (\tilde{q}_+ +\eta' q_+)/2 \right),
\label{xi_e}
\\
   \xi_{\rm p} &=& -{\rm i}\am \left(
   {\mpr\over\mH} (\eta'-\eta)K_-/\hbar - (\tilde{q}_- -\eta' q_-)/2 \right),
\label{xi_p}
\end{eqnarray}
\end{mathletters}
and
\begin{equation}
   J_{nn'}(|\xi|{\rm e}^{{\rm i}\varphi}) = 
   {\rm e}^{{\rm i}(n-n')\varphi} I_{nn'}(|\xi|^2),
\end{equation}
$I_{nn'}$ being a Laguerre function (Sokolov \& Ternov 1968). 

Using equations \req{m_shift}, \req{pi_pm}, \req{Pi_pm}, and \req{jj}
for $\eta'=1$, and neglecting 
an insignificant common phase, we finally arrive at the 
expression for the interaction matrix element~\req{m_expan}:
\begin{equation}
   \langle f | \hat{M} | i \rangle = 
   2\sqrt{\gamma} \sum_{n's'ns}
   \left[ e_+ M^{(-)}_{n's'ns} 
   + e_- M^{(+)}_{n's'ns} 
   + e_z M^{(z)}_{n's'ns} 
   \right],
\label{m_final}
\end{equation}
where 
\begin{mathletters}
\begin{eqnarray}
   M^{(-)}_{n's'ns} &=& - \left[ \sqrt{n}\, J_{n-1,n'}(\xi_{{\rm e},+}) 
   + \zeta_{{\rm e},-} J_{nn'}(\xi_{{\rm e},+}) \right]
   J_{n+s,n'+s'}(\xi_{{\rm p},+})\, {\cal{Z}}^{\rm (e)}_{n's'ns}
\nonumber\\
   &-&
   {\mel\over\mpr}\exp(-{\rm i}\eta\vv{r}_c \vv{q})
   \left[ \sqrt{n+s+1}\, J_{n+s+1,n'+s'}(\xi_{{\rm p},-})
   - \zeta_{{\rm p},-} J_{n+s,n'+s'}(\xi_{{\rm p},-}) 
   \right]
\nonumber\\
   &\times&
   J_{nn'}(\xi_{{\rm e},-})\, {\cal{Z}}^{\rm (p)}_{n's'ns},
\label{m-}\\
   M^{(+)}_{n's'ns} &=& \left[ \sqrt{n+1}\, J_{n+1,n'}(\xi_{{\rm e},+}) 
   - \zeta_{{\rm e},+} J_{nn'}(\xi_{{\rm e},+}) \right]
   J_{n+s,n'+s'}(\xi_{{\rm p},+})\, {\cal{Z}}^{\rm (e)}_{n's'ns}
\nonumber\\
   &+&
   {\mel\over\mpr}\exp(-{\rm i}\eta\vv{r}_c \vv{q})
   \left[ \sqrt{n+s}\, J_{n+s+1,n'+s'}(\xi_{{\rm p},-})
   + \zeta_{{\rm p},+} J_{n+s,n'+s'}(\xi_{{\rm p},-}) 
   \right]
\nonumber\\
   &\times&
   J_{nn'}(\xi_{{\rm e},-})\, {\cal{Z}}^{\rm (p)}_{n's'ns},
\label{m+}\\
   M^{(z)}_{n's'ns} &=& 
    \left[ \widetilde{\cal{Z}}^{\rm (e)}_{n's'ns} 
   + {{\rm i}\over 2} \am q_z {\cal{Z}}^{\rm (e)}_{n's'ns}
   \right]
   J_{nn'}(\xi_{{\rm e},+}) J_{n+s,n'+s'}(\xi_{{\rm p},+}) 
\nonumber\\
   &+&
   {\mel\over\mpr}\exp(-{\rm i}\eta\vv{r}_c \vv{q})
   \left[ \widetilde{\cal{Z}}^{\rm (p)}_{n's'ns} 
   - {{\rm i}\over 2} \am q_z {\cal{Z}}^{\rm (p)}_{n's'ns}
   \right]
   J_{nn'}(\xi_{{\rm e},-}) J_{n+s,n'+s'}(\xi_{{\rm p},-}) 
\label{mz}
\end{eqnarray}
\end{mathletters}
In these equations 
\begin{mathletters}
\begin{eqnarray}
   \zeta_{{\rm e},+} &=& - {{\rm i}\am \over\hbar} \left(
   {\mel\over\mH} (1-\eta) K_+ + \hbar q_+ \right),
~~~
   \zeta_{{\rm e},-} = - {{\rm i}\am \over\hbar} 
   {\mel\over\mH} (1-\eta) K_- ,
\label{zeta_e}\\
   \zeta_{{\rm p},\pm} &=& - {{\rm i}\am \over\hbar} 
   {\mpr\over\mH} (1-\eta) K_\pm ,
\label{zeta_p}   
\end{eqnarray}
\end{mathletters}
$\xi_{{\rm e},\pm}$ and $\xi_{{\rm p},\pm}$ are given 
by equations (A7) with $\eta'=1$ and $\tilde{\vv{q}}=\pm\vv{q}$,
respectively, and the factors 
\begin{mathletters}
\begin{eqnarray}
   {\cal{Z}}^{\rm (e)}_{n's'ns} = 
   \langle n's',f | {\rm e}^{{\rm i}q_z z} | ns\eta,i \rangle_\| ,
&~&
   {\cal{Z}}^{\rm (p)}_{n's'ns} 
   = 
   \langle n's',f | ns\eta,i \rangle_\| ,
\label{Z}\\
   \widetilde{\cal{Z}}^{\rm (e)}_{n's'ns} = 
   \langle n's',f | {\rm e}^{{\rm i}q_z z} 
   \am {\partial\over\partial z} | ns\eta,i \rangle_\| ,
&&
   \widetilde{\cal{Z}}^{\rm (p)}_{n's'ns} 
   = 
   \langle n's',f | 
   \am {\partial\over\partial z} | ns\eta,i \rangle_\|
\label{Z_tilde}
\end{eqnarray}
\end{mathletters}
are the longitudinal matrix elements subject to numerical evaluation. 

We have presented the result for $\eta'=1$. 
For arbitrary $\eta'$, the matrix elements 
are obtained in analytic form using equation (A15) of BP94 
and our equations (A6), (A7) 
with $\tilde{\vv{q}}=[2\mel + \eta'(\mpr-\mel)]\vv{q}/\mH$ 
and $\tilde{\vv{q}}=-[2\mpr + \eta'(\mel-\mpr)]\vv{q}/\mH$. 
The result is a straightforward 
generalization of equations (A10). 

} 

\newpage




\end{document}